\documentclass[a4paper,aps,pre,superscriptaddress,floatfix,nofootinbib,twocolumn,bibtex]{revtex4}

\usepackage{amsmath}
\usepackage{amssymb}

\usepackage[pdftex]{graphicx}

\usepackage{color}

\begin{document}

\title{A reverse engineering approach to the suppression of citation biases reveals universal properties of citation distributions}

\author{Filippo Radicchi}\email{f.radicchi@gmail.com}
\affiliation{Departament d'Enginyeria Quimica, Universitat Rovira i Virgili, Av. Paisos Catalans 26, 43007 Tarragona, Catalunya, Spain}
\affiliation{Howard Hughes Medical Institute (HHMI), Northwestern University, Evanston, Illinois 60208 USA}
\affiliation{Department of Chemical and Biological Engineering, Northwestern University, Evanston, Illinois 60208 USA}

\author{Claudio Castellano}
\affiliation{Istituto dei Sistemi Complessi
(ISC-CNR), Via dei Taurini 19, I-00185 Roma, Italy, and
Dipartimento di Fisica, ``Sapienza'' Universit\`a di Roma, P.le
A. Moro 2, I-00185 Roma, Italy}

\begin{abstract}
\noindent The large amount of information contained in bibliographic
databases has recently boosted the use of citations, and other indicators
based on citation numbers, as tools for the quantitative assessment
of scientific research.
Citations counts are often interpreted as proxies for the scientific
influence of papers, journals, scholars, and institutions.
However, a rigorous and scientifically grounded methodology for a
correct use of citation counts is still missing. In particular,
cross-disciplinary comparisons in terms of raw citation
counts systematically favors scientific
disciplines with higher citation and publication rates.  
Here we perform an exhaustive study of the citation patterns
of millions of papers, and derive a simple transformation of
citation counts
able to suppress the disproportionate citation counts
among scientific domains. We find that 
the transformation is well described by a power-law function, and that
the parameter values of the transformation are typical
features of each scientific discipline.
Universal properties of citation patterns
descend therefore from the fact that
citation distributions for papers in a specific field
are all part of the same family of univariate distributions. 
\end{abstract}

\maketitle

\section{Introduction}
\noindent The use of bibliographic databases
plays a practical, and crucial, role in modern science. 
Citations between scientific
publications are in fact commonly used
as quantitative indicators for the importance of scientific papers,
as proxies for the influence of publications in the scientific
community. General criticisms to the use of
citation counts have been made~\cite{macroberts89,macroberts96,adler09},
and the real meaning of a citation
between papers can be very different and context dependent~\cite{bornmann08a}.
Nevertheless, a citation can be viewed as a tangible acknowledgment
of the citing paper to the cited one.
Thus, the more citations a paper has accumulated,
the more influential the paper can be considered for
its own scientific community of reference. The same unit of measure
(i.e., a citation) is commonly used as the basis
for the quantitative evaluation of
individual scholars~\cite{hirsch05,egghe06}, journals~\cite{garfield06}, 
departments~\cite{davis84}, universities and institutions~\cite{kinney07},
and even entire countries~\cite{king04}. 
Especially at the level of individual scientists, 
numerical indicators based on citation counts are evaluation tools
of fundamental importance for decisions about hiring~\cite{bornmann06}
and/or grant awards~\cite{bornmann08b}. 

\

\noindent
As a matter of fact, citation practice is widespread, 
still basic properties of citation patterns are not completely clear. 
For example, we know that citations
are broadly distributed, but, we do not know
the exact functional form of citation distributions. 
In his seminal paper,
de Solla Price proposed a power-law model for explaining
how papers accumulate citations~\cite{desolla65}. 
However, more recent studies indicate several, sometimes
very different, possibilities: power-laws~\cite{redner98, seglen99},
stretched exponentials~\cite{lahe98, wallace09},
log-normals~\cite{stringer08, radicchi08, stringer10}, and
modified Bessel functions~\cite{vanraan01}.
\\
At the same time, it is common practice to attribute the same value
to each citation, in spite of the fact that citation counts
strongly depend on the field~\cite{hamilton91}.
For example, a  paper in mathematics typically gets less citations
than a paper in molecular biology. There are in fact large variations
among scientific communities, mostly related
to the different citation habits of each community.
Such disproportions show up in
the typical values of the most common bibliometric indicators
based on raw citation counts.
The most influential journal in mathematics, {\it Annals
of Mathematics}, has impact factor~\cite{garfield06}
roughly equal to $4$ according to the $2009$
edition of the Journal Citation Reports (JCR)
database~\cite{jcr2009}, while its counterpart in molecular biology, 
{\it Cell}, has impact factor $32$, eight times larger.
 Similarly, there are several chemists with 
$h$-index~\cite{hirsch05} larger 
than $150$~\cite{chem}, while for a computer scientist
it is very hard to have an $h$-index larger 
than $100$~\cite{compsci}. Notice
that the values of the $h$-index for chemists 
have been calculated in $2007$, while those for
computer scientists in $2011$. For the same
year of reference, we should expect that the 
difference is even larger than what reported here.
Such disproportions in citation counts make the use of raw
citation numbers very precarious 
 in many cases and call for alternative, more fair, measures.
It is important to stress that in this paper we denote as ``bias''
the the systematic error that is introduced when using raw citation
numbers to compare papers belonging to different fields.
With this term we do not indicate any prejudice, nor we make any
claim about the causes of the field dependence empirically observed.
\

\noindent
Although methods based on percentile ranks
have been recently considered~\cite{leydes11, bornmann11},
the traditional approach to the suppression
of field-dependence in citation counts is based on normalized indicators.
The raw number of citations is divided by a discipline dependent factor,
and the aim of this linear transformation is to suppress
eventual disproportions among the citation patterns
of different research fields. Various methods
have been proposed using 
this kind of approach~\cite{schubert96,vinkler96,vinkler03,zitt05,leydes10}.
In this context, particularly relevant is the study
performed in~\cite{radicchi08} (based on the
the relative indicator originally developed in~\cite{lundberg07}), 
where citation distributions of different 
scientific disciplines are shown to have the same functional form, 
differing only for a single scaling factor (the
average number of citations received by papers
within each scientific discipline).
The study is, however, limited to a small number of papers and
scientific disciplines, and therefore not conclusive.
The same approach of~\cite{radicchi08} has also been used
for more refined classification
of publications in physics~\cite{radicchi11} and
chemistry~\cite{bornmann09}, showing
in general a good agreement with
the previous claim of~\cite{radicchi08}. More recently,
Albarr\'an {\it et al.}~\cite{albarran11} and 
Waltman {\it et al.}~\cite{waltman11} have analyzed much larger
datasets of scientific publications, and showed
that the result of~\cite{radicchi08} holds for many
but not for all scientific disciplines. These studies
cast some doubts on the validity of the results in~\cite{radicchi08},
but, on the other hand, do not propose any alternative method for bias
suppression.

\

\noindent
Here, we perform an exhaustive analysis of about $3$ millions of papers
published in six different years (spanning
almost $30$ years of scientific production)
and in more than $8,000$
journals listed in the Web Of Science (WOS)~\cite{wos} database. 
We use the classification of journals in subject-categories ($172$ in total)
as defined in the $2009$ edition of the Journal
Citation Reports (JCR) database~\cite{jcr2009}, and systematically
study the patterns of citations received by
papers within single subject-categories.
Despite some journals cover a rather broad range of topics, 
a subject-category is a relatively accurate
classification of the general content of a journal. 
Examples of JCR subject-categories are ``Mathematics'', ``Reproductive
Biology'' and ``Physics, Condensed Matter''. 
Subject-categories can be
considered as good approximations for scientific disciplines.
\\
We propose a transformation of raw citations numbers such that
the distributions of transformed citation counts are the same
for all subject-categories.
We study the properties
of this transformation and find strong regularities among
scientific disciplines. The transformation is almost
linear for the majority of the subject-categories. 
Exceptions to this rule are present, but, in general, we find
that all citation distributions are part
the same family of univariate distributions.
In particular, the rescaling considered in~\cite{radicchi08}, 
despite not strictly correct, is
a very good approximation of the transformation
able to make citation counts not depending on the
scientific domain.

\section{Results}
\subsection{Modeling citation distributions}
\noindent
For the same year of publication, the raw citation patterns
of single subject-categories may be very different.
Variations are a consequence of
different publication and citation habits among scientific disciplines. 
In Fig.~1 for example, we plot the cumulative distributions
of citations received by papers published
in journals belonging to three different subject-categories.
The shape of the three cumulative distributions is not exactly
the same, and the difference is not accounted for by a
single scaling factor~\cite{radicchi08}.
Dividing raw citation counts by a scaling factor (e.g., the
average number of citations of the subject-category) 
would in fact correspond, in the logarithmic scale,
to a horizontal rigid translation of the cumulative distribution. 
However, as Fig.~1 shows, this linear transformation is not
sufficient to make all cumulative distributions coincide.
By looking at the figure, the cumulative  distributions of the raw citation 
counts for papers published in journals within the subject-categories
``Computer science, software engineering'' and ``Genetics \& heredity''
have a pretty similar shape, and thus the possibility
to obtain a good collapse of the curves by simply rescaling
citation counts seems reasonable. Conversely, the cumulative
distribution of the citations received by papers published in journals of 
the subject-category ``Agronomy'' has a  different shape. The curve 
bends down faster than the curves corresponding 
to the other two subject-categories. In this case, a linear
transformation of citation counts would hardly help  
to make this curve coincide with the others.
Making citation counts independent of the
subject-categories seems therefore not possible with
the use of linear transformations,
because the difference between citation distributions
of different subject-categories is not only due 
to a single scaling factor.

\

\noindent
In order to make further progress, here we
invert the approach to the problem.
We know that citation patterns of single subject-categories
may be different, but we do not know how to transform citation counts in order
to make them similar. We implement therefore a mapping able to make
all cumulative distributions coincide, and
study the properties of this transformation. 
We use a sort of ``reverse engineering'' approach:
instead of introducing a transformation and checking whether it works,
here we impose that the transformation must work and from this assumption
we derive its precise form.
\\
The idea is pretty simple and straightforward. 
We use as curve of reference 
the cumulative distribution $P\left(\geq c\right)$ of
raw citation counts $c$ obtained
by aggregating together all subject-categories (see Fig.~1).
\begin{figure}[!htb]
\includegraphics[width=0.48\textwidth]{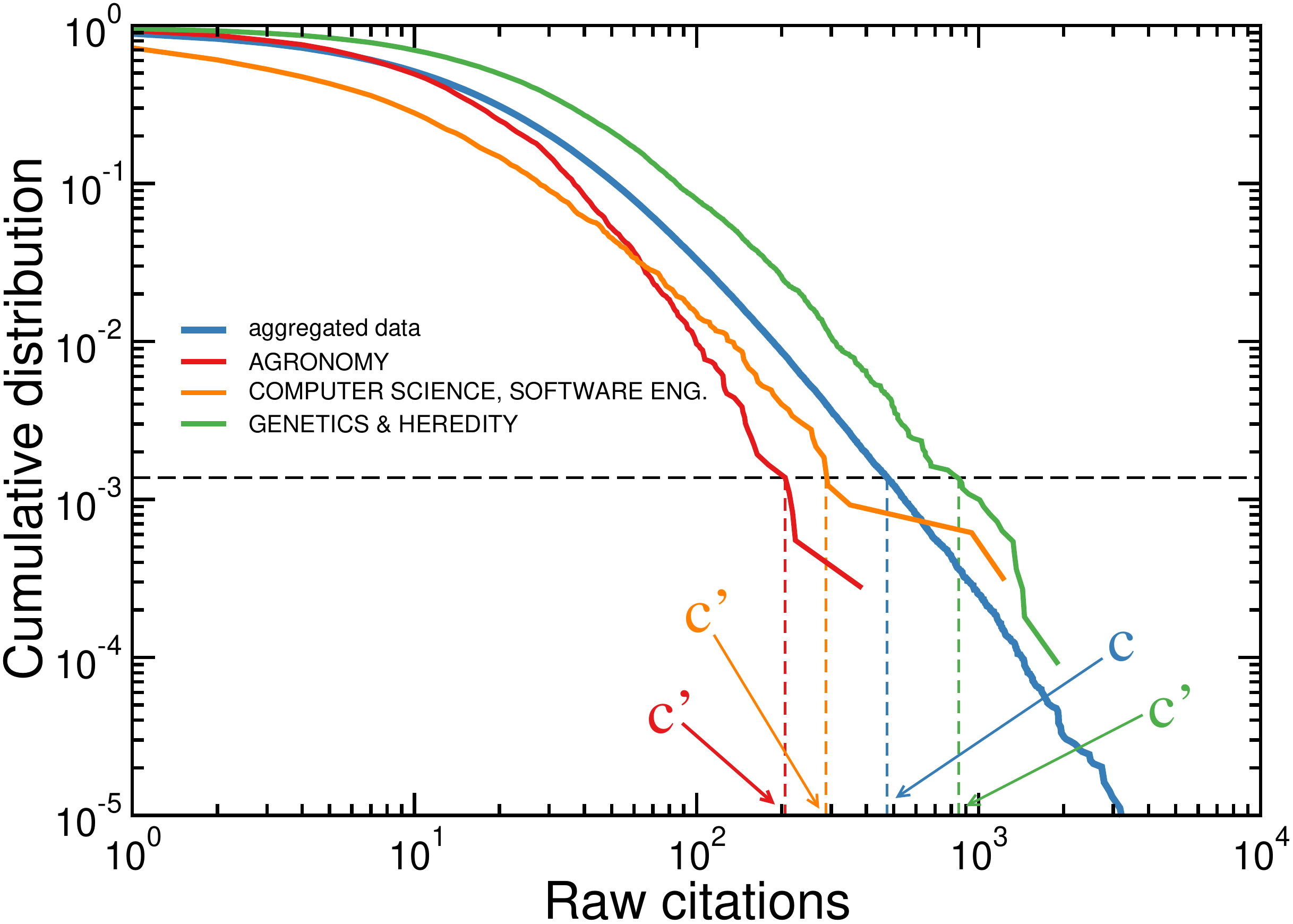}
\caption{Cumulative distribution of raw citation counts
for papers published in $1999$.
The blue curve is calculated by aggregating
all papers of all subject-categories
 (average number
of citations $\langle c \rangle = 21.97$). The red curve,
the orange curve and the green curve
are calculated by considering only papers within
the subject-categories ``Agronomy'' ($\langle c' \rangle = 15.62$),
``Computer science, software engineering'' ($\langle c' \rangle = 11.57$)
and ``Genetics \& heredity'' ($\langle c' \rangle = 38.87$), respectively.
The figure illustrates the mapping of $c'$ into $c$.
Citation counts $c'$ of single subject-categories
are matched with the value of $c$ which
corresponds to same value of the cumulative distributions.
}
\end{figure}
The choice of the curve of reference is in principle arbitrary, and
affects the explicit form of the transformation.
The use of the aggregated dataset as reference seems, however,
a very reasonable choice because it does not require the introduction of 
any parameter. In general, other
choices for the reference curve are possible,
but the only important constraints are (i) using the same system of 
reference for all subject-categories and (ii) producing
a mapping that preserves the natural order of citation
counts within the same subject-category. 
We then focus on a specific subject-category $g$, and consider
the cumulative distribution $P_g\left(\geq c'\right)$ 
of the raw citations $c'$ received by papers published 
in journals within subject-category $g$. 
To each value of $c'$, we associate a single value of
$c$ in the system of reference, where $c$ is determined as the value
for which $P_g\left(\geq c'\right)=P\left(\geq c\right)$.
In practice, we implement the mapping 
by sorting in ascending order all citation counts of the $N$
papers present in the aggregated dataset, and then by associating to 
each different value of $c'$, in the
dataset of subject-category $g$, the value of $c$ that appears in 
the $n$-th position of the sorted list, 
with $n$ equal to the integer value closest
to $N\; P_g\left(\geq c'\right)$.
In this procedure, different values of $c'$
may correspond to the same value of $c$. Such event is
more likely to happen for low values of $c'$, while,
for large values of $c'$, the mapping is always unique (see Fig.~2).
\begin{figure}[!htb]
\includegraphics[width=0.48\textwidth]{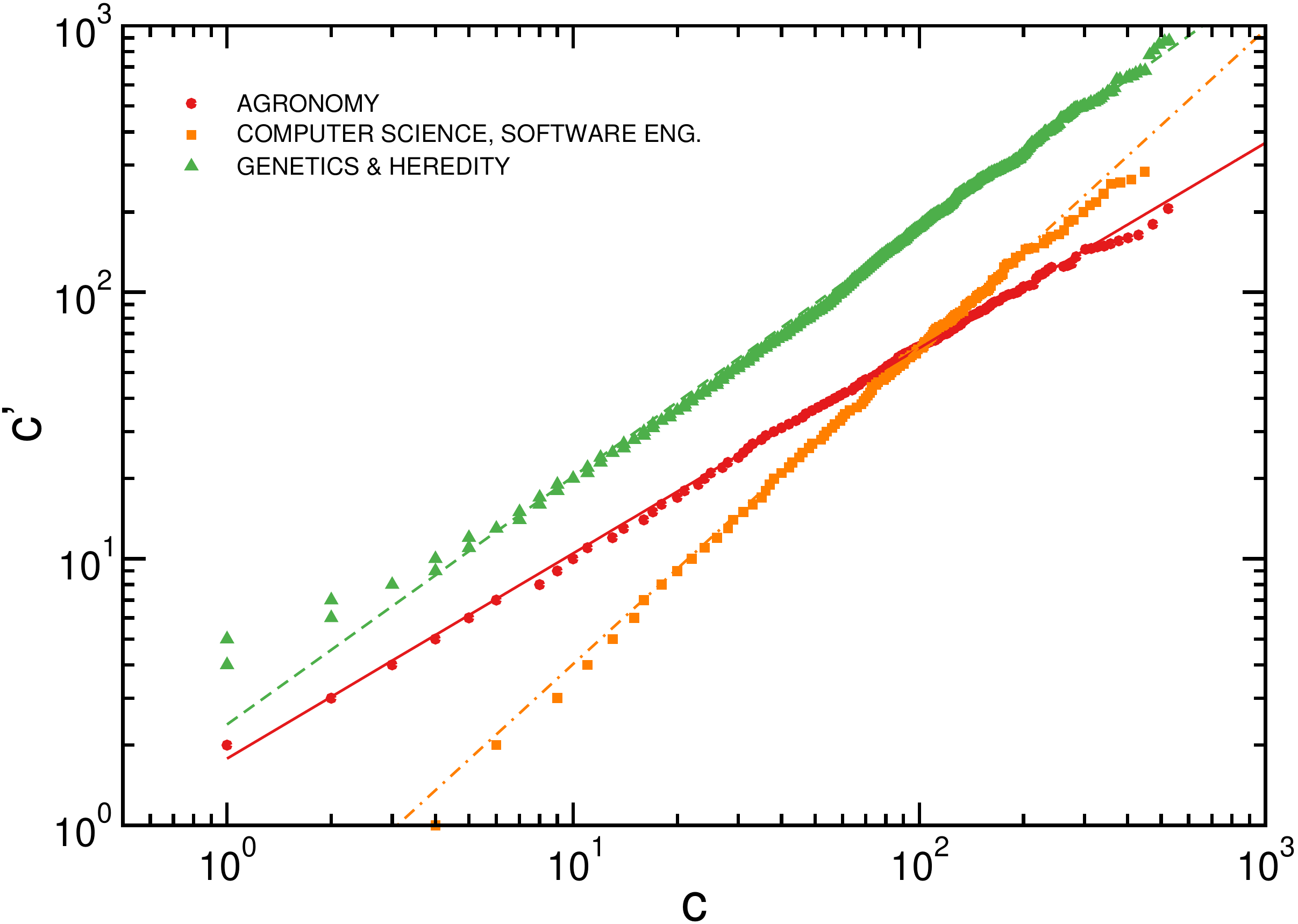}
\caption{Transformation of citation counts.
Citations within single subject-categories.
 $c'$ are plotted against citation counts
of the aggregated data $c$. The quantities $c'$ and $c$
are related by a power-law relation (Eq.~\ref{eq1}).
Different subject-categories have different values
of the transformation factor $a$ and the transformation
exponent $\alpha$. The best estimates
of $a$ and $\alpha$ for the subject-categories
considered in this figure (the
same subject-categories as those appearing in Fig.~1) are:
$a=1.78 \pm 0.02$ and  $\alpha = 0.77 \pm 0.01$ 
for ``Agronomy'', $a=0.26 \pm 0.01$ and $\alpha=1.19 \pm 0.01$ for ``Computer
science, software engineering'', $a=2.39 \pm 0.04$ 
and $\alpha=0.93 \pm 0.01$ for ``Genetics \& heredity''.
The results of the complete analysis for all subject-categories and years
of publication are reported in the Supporting
Information S2-S7.
}
\end{figure}

\

\noindent
The plot $c'$ {\it vs.} $c$ is 
equivalent to a quantile-quantile ($Q-Q$) plot,
a graphical non-parametric method generally used for comparing
two probability distributions~\cite{wilk68}.
If the comparison is made between two samples of randomly and identically
distributed variates, all points in the corresponding $Q-Q$ plot, 
should approximately lay on the line $y=x$.
If the difference between the two samples
is just a scaling factor $a$, then all points
in the $Q-Q$ plot should instead lay on the line $y=a\,x$.
Very interestingly in the case of citation distributions, 
we empirically find that the relation between $c'$ 
and $c$ can be described by a power-law function
\begin{equation}
c' = a \, c^\alpha \;\;,
\label{eq1}
\end{equation}
where $a$ and $\alpha$ are respectively the pre-factor and the
exponent of the mapping (see Fig.~2).
The functional 
form of Eq.~\ref{eq1} holds for virtually all subject-categories 
and all publication years considered in this study 
(see Supporting Information S2-S7). Few exceptions are present, 
the most noticeable represented by the 
hybrid subject-category ``Multidisciplinary sciences''.

\

\noindent
The citation distributions of the subject-categories 
for which Eq.~\ref{eq1} holds
are univariate distributions belonging to the same 
log-location-scale family~\cite{stat82}. 
A log-location-scale family of distributions 
is a class of distributions $p\left(\log{x};\theta,\delta\right)$ 
of continuous variables $x$ that can be rewritten
in terms of the same reference distribution $r\left(\cdot\right)$ as
$p\left(\log{x};\theta,\delta\right)= \delta^{-1} \, r\left(\left(\log{x}-\theta \right)/\delta\right)$, for any choice of
the location parameter $-\infty < \theta < \infty$ and the scale parameter
$\delta > 0$~\cite{stat00}.
Citation distributions are
defined for discrete variables, but
still according to Eq.~\ref{eq1} we can write 
$P_g\left(\geq c'\right) = P\left(\geq a c^\alpha\right)$,
where $a$ and $\alpha$ respectively represent the log-location and 
the log-scale parameters.
In few words, our empirical finding tells us that citation distributions 
are part of the same log-location-scale family of discrete distributions.
Weibull and log-normal distributions are well known 
log-location-scale families.

\subsection{Cumulative distribution of transformed citations}
\noindent
By definition, the transformation $c' \to c$
maps the cumulative distribution on top of the 
cumulative distribution of reference 
(i.e., the one calculated for the aggregated data).
Therefore, if the same transformation is 
applied to the citation numbers
of all subject-categories, all cumulative distributions concide,
providing a systematic deletion of
differences present in the citation patterns. 
Eq.~\ref{eq1} tells us that the mapping $c' \to c$ is
simple. The citations $c'$
received by papers published in journals within a specific
subject-category can be simply transformed as
\begin{equation}
c' \to c = \left(\frac{c'}{a}\right)^{\frac{1}{\alpha}} \;\; ,
\label{eq2}
\end{equation}
if we want to make all citation distributions of single
subject-categories coincide with the cumulative
distribution of reference. Fig.~3 shows the
cumulative distributions resulting after the
application of Eq.~\ref{eq2}.
The cumulative distributions of the transformed
citation counts are very similar. Small deviations
are still visible at low values of the transformed
citation counts, when the discreteness 
of citation numbers become more important.
\begin{figure}[!htb]
\includegraphics[width=0.48\textwidth]{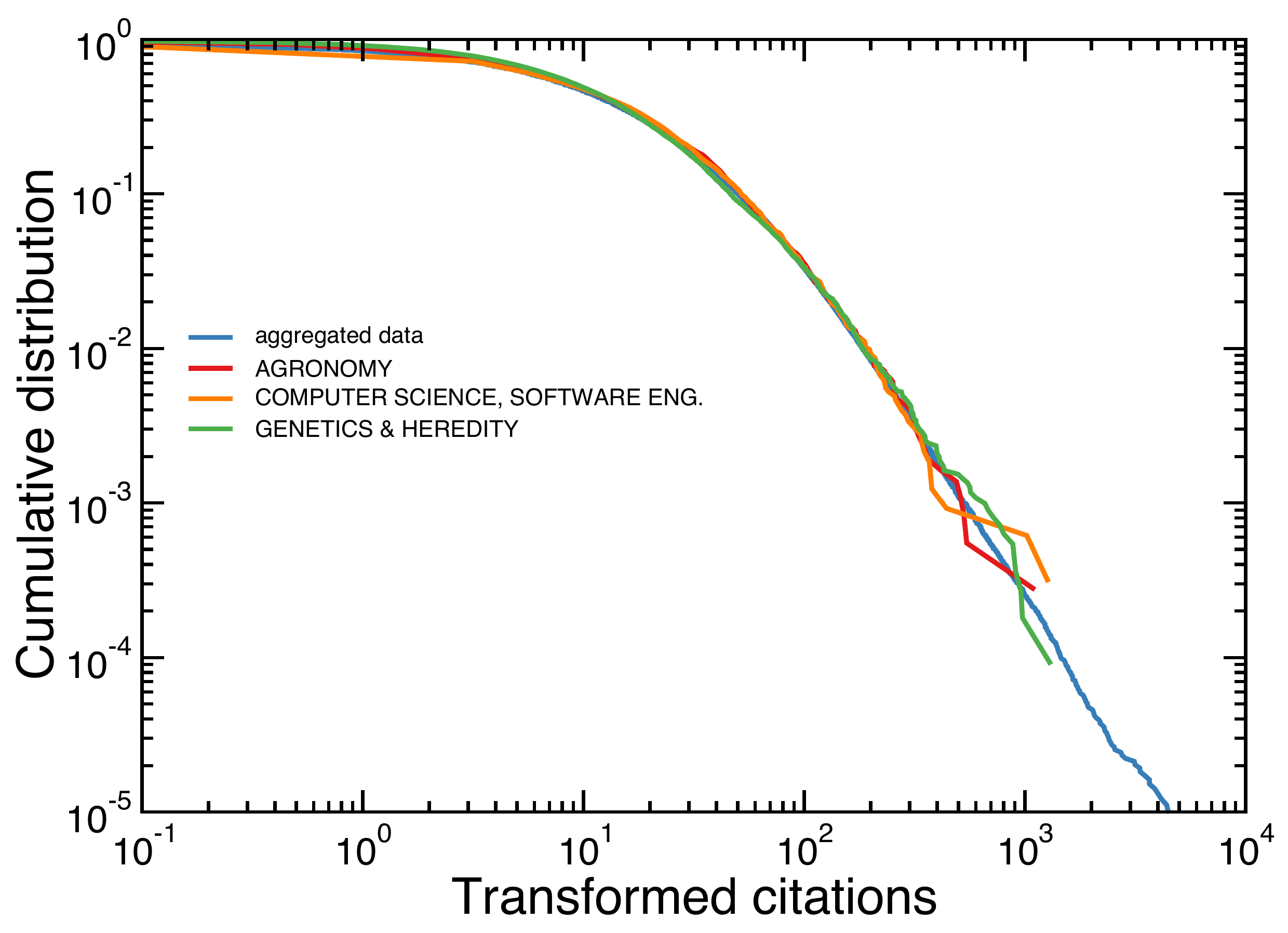}
\caption{Cumulative distribution of the transformed
citation counts.
When raw citation numbers are transformed
according to Eq.~\ref{eq2}, the cumulative distributions
of different subject-categories become very
similar. All citation distributions are mapped
on top of the cumulative
distribution obtained by aggregating all subject-categories
together (the common reference curve in the transformation). We consider here 
the same subject-categories as those considered in Figs.~1 
and 2. The complete analysis of all subject-categories
and years of publication is reported
in the Supporting Information S2-S7.
}
\end{figure}

\subsection{Quantitative test of bias suppression}
\noindent
The fact that all cumulative distributions 
of transformed citation counts coincide seems able to
place all subject-categories on the same footing:
when raw citations are transformed
according to Eq.~\ref{eq2}, the fraction of papers 
with a given value of the transformed citation counts 
is almost the same for all subject-categories. 
To quantitatively assess such a qualitative result, 
we perform an additional test.
The aim of the transformation of Eq.~\ref{eq2} is
to suppress inevitable 
biases in raw citation counts
among subject-categories, and thus we
compare our results with the outcome expected
in the absence of biases.
\\
The situation can be modeled in the following terms.
We aggregate all papers of all subject-categories
together, and extract the top $v\%$ 
of
publications according to the
value of their transformed citations.
We then compute the proportion of papers in each
subject-category that are part of the top $v\%$ 
Assuming all cumulative distributions to be the same, we
expect these proportions to have values close to $v/100$. 
However, since the number of papers in each subject-category
is finite, the proportions of papers
belonging to the top $v\%$ 
are affected by fluctuations,
which can be precisely computed (see the Methods section for details).
By checking if the outcome of our selection process
is compatible with the results expected assuming
a random and unbiased selection process,
we test whether we have effectively removed citation biases.
\\
\begin{figure}[!htb]
\includegraphics[width=0.48\textwidth]{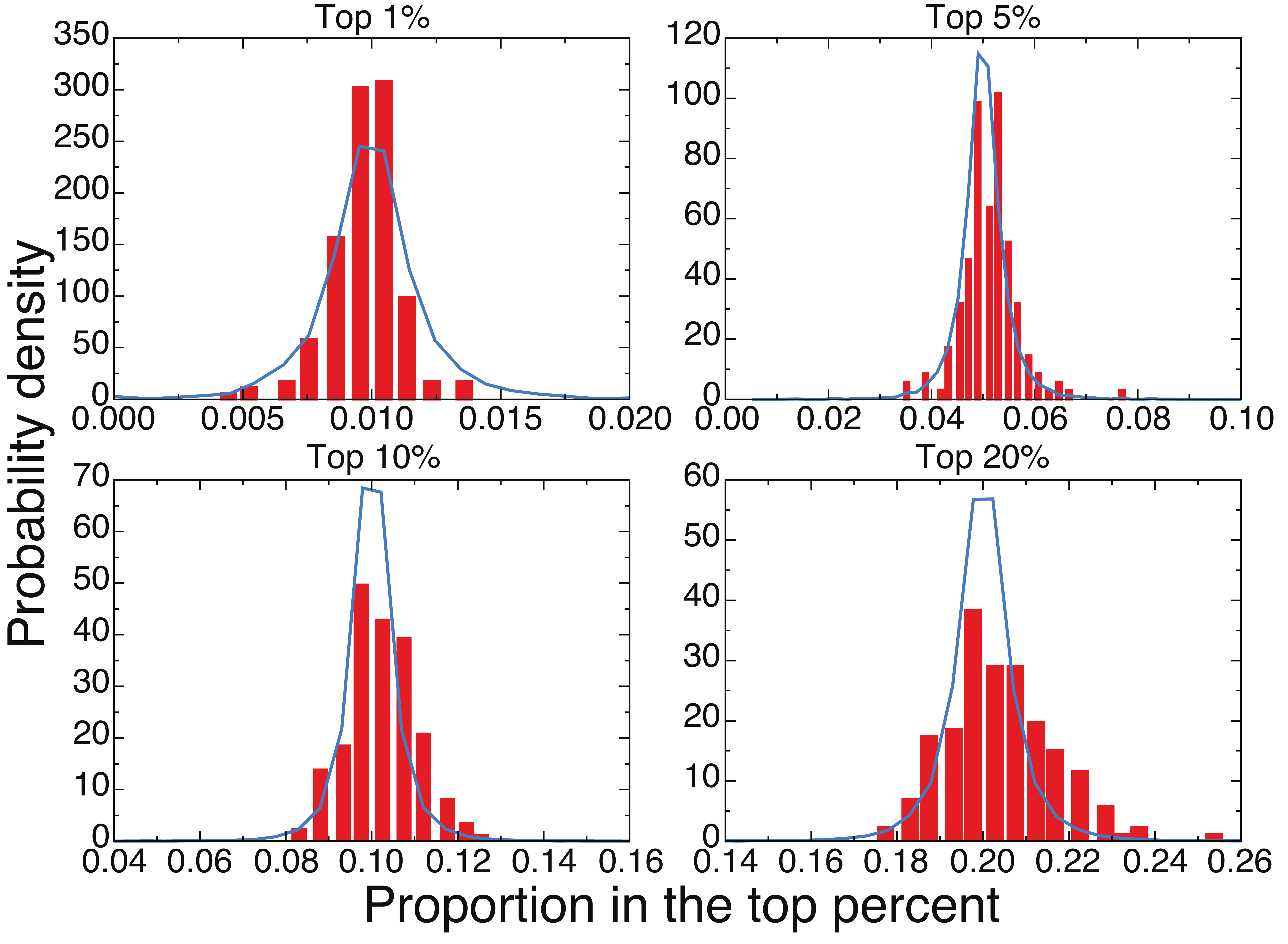}
\caption{Comparison between expected and observed
proportions of top cited papers.
Probability density function of
the proportion of papers belonging to a particular
subject-category and that are part of the top $v\%$ 
of papers in the aggregated dataset. 
Red boxes are computed on real data, while
blue curves represent the
density distributions valid for unbiased
selection processes.
We consider different
values of $v$: 
$1$, $5$, $10$ and $20$.
These results refer to papers published in $1999$.
}
\end{figure}
\begin{figure}[!htb]
\includegraphics[width=0.48\textwidth]{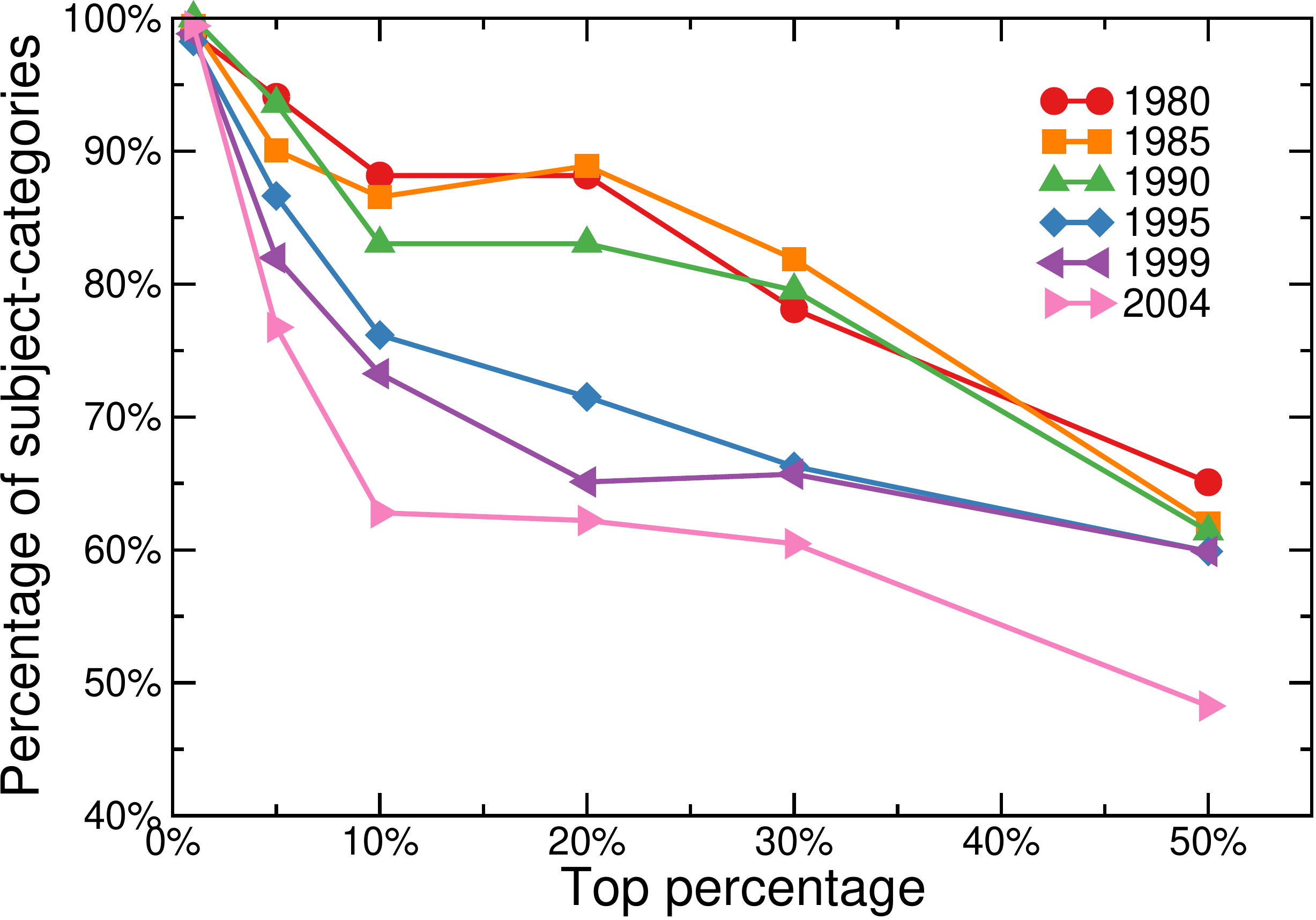}
\caption{Effectiveness of the proposed
normalization technique.
Percentage of subject-categories whose proportion values,
after normalization, fall into the 95\% confidence interval 
of values predicted in our null model. Percentage values
are plotted as functions of the percentage $v\%$ of top
papers considered in the analysis.
We plot separate curves
for different publication years.
}
\end{figure}
The results of this analysis are reported in Fig.~4 
for papers published in $1999$,
and in the Supporting Information S1-S7 
for other publication years.
In general, the transformation of Eq.~\ref{eq2} produces, 
for all years of publication,
results that are consistent with an unbiased
selection process, if $v \leq 10$ (see Fig.~5). 
For the most
relevant part of the curve (i.e., highly cited
papers), the simple transformation of Eq.~\ref{eq2}
effectively removes systematic
differences in citation patterns
among subject-categories. Conversely, for higher values
of $v$, 
the discreteness of citation numbers becomes
more relevant, the power-law mapping
of Eq.~\ref{eq1} becomes less descriptive, 
and the distribution of the proportion
of top $v\%$ 
papers measured for real data, despite
still centered around the expected value, is wider than
expected. The results are even better for papers published
before year $1995$ because
the comparison between observed and expected proportions
of papers in the top $v\%$ 
is very good up to $v=30$. 
The reason could be due to a higher stability
of citation patterns for all subject-categories, 
since all papers have had more than $15$ years
to accumulate citations~\cite{stringer08}.

\subsection{Values of the transformation parameters}
\noindent
The values of the transformation factor $a$ and
the transformation exponent $\alpha$ for the same
subject-category are pretty stable when measured
over different years of publication.
In particular, the value of $\alpha$ 
is very robust, suggesting that the shape
of the cumulative distribution of
single subject-categories does not vary
with time. For example, over
a span of almost $30$ years, the values $\alpha$ for
the subject-category ``Agronomy'' range
in the interval $\left[0.74,0.82\right]$,
for ``Computer science, software engineering'' 
range in the interval $\left[1.04, 1.32 \right]$,
and for ``Genetics \& heredity'' range in the 
interval $\left[0.86, 0.93 \right]$.
Tables
reporting the complete results for all subject-categories
and publication years can be found in the
Supporting Information S2-S7.
The density distribution
of the transformation exponents is peaked
around $0.85$ which means that the shape
of the distributions is in the majority
of the cases the same and
the only difference is a scaling factor (see inset of Fig.~6 and Supporting
Information S8).
\\
\begin{figure}[!htb]
\includegraphics[width=0.48\textwidth]{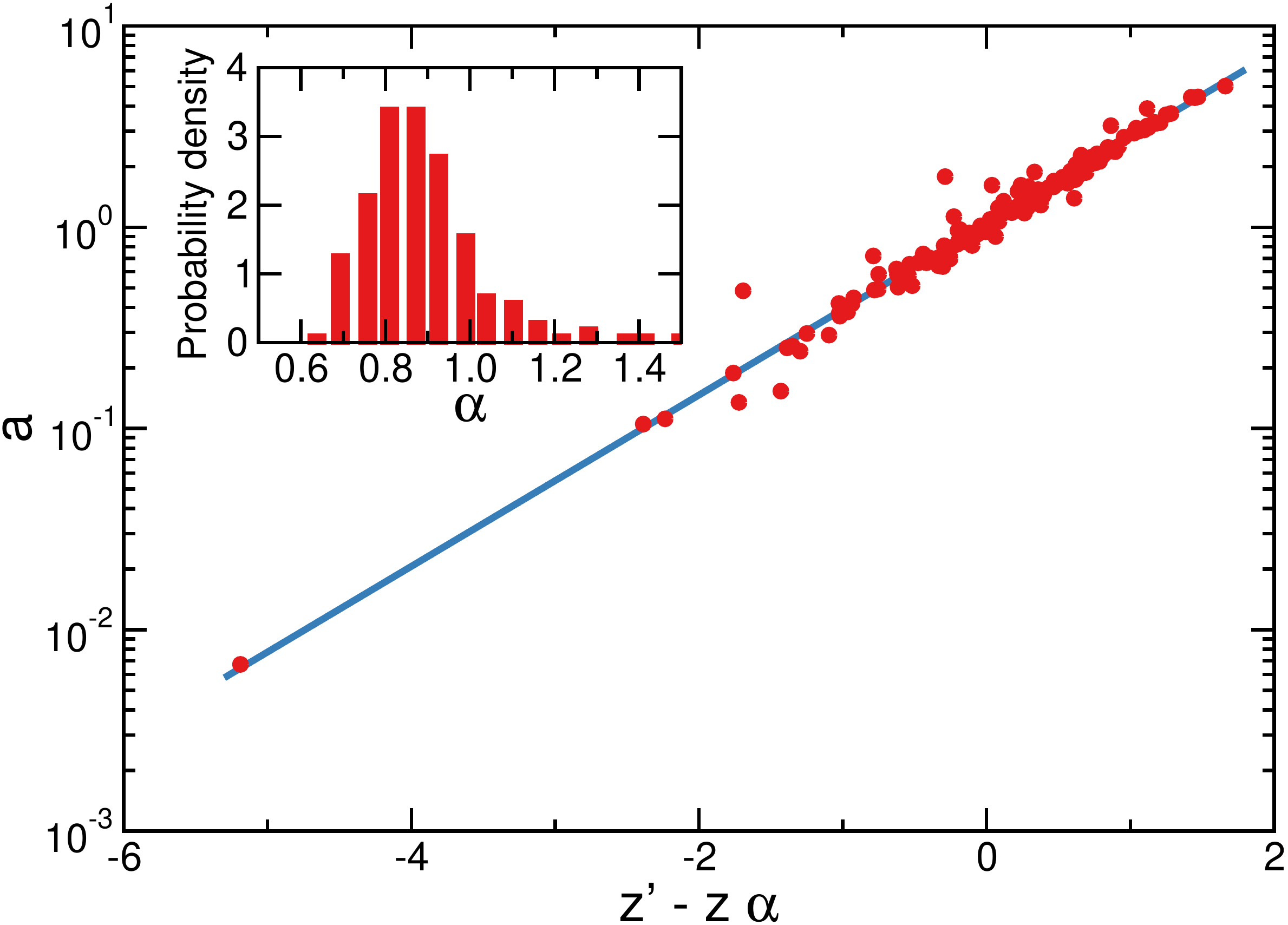}
\caption{Properties of the transformation
parameters.
In the inset, we
report the density distribution
of the transformation exponents $\alpha$
calculated for all subject-categories.
In the main plot, we show the relation
between the transformation exponent $\alpha$, 
the transformation factor $a$, and the parameters
$z'$ and $z$
for the same data points as those appearing in the inset.
The relation between the various quantities
is fitted by the function $a=e^{r + t \left(z' -z \alpha\right)}$, 
with $r=0.04 \pm 0.01$ and $t=0.98 \pm 0.01$ (blue line).
Both plots have been obtained
by analyzing papers published in $1999$, but the same results
are valid also for different years of publications as 
shown in Figs.~S115 and S116.
}
\end{figure}

\noindent
Moreover, the transformation factor $a$ and the transformation exponent
$\alpha$ are related.
Let us consider what happens for log-normal distributions.
A log-normal distribution is given by
$P\left(x\right) = \frac{1}{\sqrt{2 \pi} \, s \, x} \; e^{-\left[\log{\left(x\right)}-z\right]^2 \left/ \left(2 s^2\right) \right.}$, 
where $z$ and $s$ are the parameters of the distribution.
The parameters $z$ and $s$ are related to
the mean $\langle x \rangle$ and variance $\sigma$ of
the distribution:
$z= \log{\left(\langle x \rangle\right)} - s^2/2$ 
and $s^2=\log{\left[\left(\sigma/\langle x \rangle\right)^2+1\right]}$.
A $Q-Q$ plot between two log-normal distributions
with parameters $z'$ and $s'$, and $z$ and $s$, respectively, shows
a perfect power-law scaling as the one given by
Eq.~\ref{eq1}. In this case, $a$ and $\alpha$ are related to the parameters 
of the distributions by
\begin{equation}
a = e^{z' - z \alpha} \qquad \textrm{ and } \qquad \alpha = \frac{s'}{s} \;.
\label{eq:log-normal}
\end{equation} 
We checked whether Eq.~\ref{eq:log-normal} is valid also
in the case of the citation distributions considered here.
In Fig.~6, we show the results obtained for publication year $1999$, while 
the plots for the other publication years are reported in Supporting Information S8. 
In general for citation distributions, Eq.~\ref{eq:log-normal} 
should be generalized to
$a=e^{r + t \left(z' -z \alpha\right)}$, with small but non vanishing 
values of $r$ and values of $t$ slightly different from one.
We conclude that the citations for single
subject-categories are distributed almost log-normally and this reflects
in the values of transformation parameters.
\\
The $Q-Q$ plot between two log-normal distributions helps also
understanding why the typical values
of $\alpha$ are generally smaller than one (inset of Fig.~6 and Supporting Information S8).
According to our choice, the reference
distribution is given by the aggregation of all subject-categories, 
and this means that the variance of the resulting distribution is mainly
determined by those of the subject-categories with higher variances. 
For the majority of the subject-categories we have $s' < s$, that is
$\alpha<1$.

\section{Discussion}
\noindent The practical importance of citation counts in modern science is
substantial, and growing.
Citation numbers (or numerical indicators derived from them) 
are commonly used as basic units of measure for the scientific
relevance
not only of papers, but also of scientists~\cite{hirsch05,egghe06}, journals~\cite{garfield06}, 
departments~\cite{davis84}, universities and institutions~\cite{kinney07},
and even entire countries~\cite{king04}.
Citations are direct measures of popularity and influence,
and the use of citation numbers is a common evaluation tool
for awarding institutional positions~\cite{bornmann06} and
grants~\cite{bornmann08b}.
Unfortunately, the direct use of raw citations
is in most of the cases misleading, especially when
applied to cross-disciplinary comparisons~\cite{hamilton91}.
Citations have different weights depending on the context
where they are used, and proper scales of measurements
are required for the formulation of objective quantitative 
criteria of assessment. Saying that a paper in biology
is more influential than a paper in mathematics,
only because the former has received a number of citations
three times larger than the latter, is incorrect.
Differences in publication and citation habits among
scientific disciplines are reflected in citation and publication counts, and
generally cause disproportions that favor disciplines
with higher publication and citation rates with respect to those
disciplines where publications and citations are created
at slower rates. In a certain sense, the situation is similar
to the comparison of the length of two streets, one long
three and the other two, but without knowing that
the length of the first is measured in kilometers while the other
in miles.

\

\noindent
Differences in citation patterns among scientific domains have been
known for a long time~\cite{hamilton91} and several attempts to the suppression 
of discipline dependent factors
in raw citation counts have been already proposed in the
past~\cite{schubert96,vinkler96,vinkler03,zitt05,radicchi08,leydes10}.
The most common methodology consists in dividing citation counts by 
a constant factor, and thus replacing raw 
with normalized citation numbers.
Each normalization procedure is, however, based on some assumption.
Scientific disciplines differ not only in citation numbers,
but also in publication numbers,
length of references and author lists, etc. A universal criterion
for the complete suppression of differences among
scientific domains probably does not exist. There are too many factors
to account for, and consequently the "philosophy" at the basis of 
a "fair" normalization procedure is subjective. 
The formulation of the so-called fractional citation count 
is, for example, based on a particular
idea of fairness~\cite{leydes10}.
Citations are normalized by assigning to each citation originated by a paper
a weight equal to the inverse of the total number of cited
references in that paper.
According to this procedure, the weight of each published
paper equals one, but disciplines with higher publication rates are
still favored when compared with disciplines with lower publication rates.

\

\noindent
In this paper, we consider a different
notion of fairness, based on the reasonable but strong assumption 
that each discipline or field of research
has the same importance for the development of
scientific knowledge. A fair numerical
indicator, based on citation numbers, must then assume
values that do not depend on the particular scientific domain
under consideration. Under this assumption,
the probability to find a paper with a given value
of the fair indicator must not depend
on the discipline of the paper, or equivalently, the distribution
of normalized indicators must be the same
for
all disciplines. It is clear that our notion
of fairness strongly depends on the
classification of papers into categories (disciplines,
fields, topics). Also, it is important to remark that
other possible definitions of fairness could be stated,
without relying on the assumption that each discipline
or research field has the same importance for scientific development.
\\
We have then proposed a simple but rigorous method
for the implementation of our notion of fairness.
We have studied the citation patterns of papers published
in more than $8,000$ scientific journals. Our analysis
covers six different years of publication, spanning
over almost $30$ years of scientific
production, and includes three millions of papers. 
We have found strong regularities
in how citations are attributed to papers dealing with
similar scientific topics of research (i.e., subject-categories).
In particular, we have introduced a simple mapping
able to transform the citation distribution of papers
published within specific subject-categories into 
the same distribution. 
Very interestingly,
the transformation turns out to be described by a power-law function,
which depends on two parameters 
(pre-factor and exponent). 
Each specific subject-category is characterized by its parameters,
which are stable over different publication years.
For the vast majority of the subject-categories, 
the power-law exponent assumes approximately the same value
suggesting that the main difference between the citation
distribution of different subject-categories
is given only by a scaling factor. 
There are, however, subject-categories for which
the transformation is not a power-law function.
In general, these are hybrid subject-categories,
as for example ``Multidisciplinary sciences'', or not so well
defined subject-categories, as for example ``Engineering, petroleum''
or ``Biodiversity conservation''. In the latter cases, the subject-categories
are not well defined because papers within these subject-categories
are also part of other broader subject-categories.
Since the classification of JCR is made at journal level, papers
published in multi-category journals are automatically attributed
to more subject-categories. In this way for example, $100\%$ of papers 
published in $1999$ in journals
within the subject-category ``Biodiversity conservation'' are also part of
``Ecology'', and $90\%$ of papers published in $1999$ within
``Engineering, petroleum'' are also part of ``Energy \& fuels''.
These observations cast some doubts regarding the
classification of JCR, which probably requires
serious revisions, especially because it seems that
the classification places on the same footing very broad 
subject-categories
and more specific ones.
Despite that, the results reported in this paper
support the claim that citation distributions
are universal, in the sense that they are all part
of the same family of univariate
distributions (i.e., a log-location-scale family~\cite{stat82,stat00}).
Each citation distribution can be obtained from the same reference
distribution with the only prescription of transforming
the logarithm of its argument with suitably chosen
location and scale parameters. 
The transformation generalizes therefore the rescaling
of~\cite{radicchi08},
that can be considered a good approximation
of the full transformation able to suppress field-dependent differences
in citation patterns.

\

\noindent
In general, all results obtained in this paper
could seem to be explained by assuming that the
citations received by papers in each subject-category
are continuous variables obeying log-normal distributions.
However, this is only approximately true. 
First, citations are, by definition, non negative discrete numbers.
Secondly, even assuming their discreteness, the
distribution of citations received
by papers within the same subject-category is not statistically
consistent with a discrete log-normal distribution.
We systematically tested this hypothesis
for all subject-categories
and publication years, and found that 
the log-normality of citation distributions 
cannot be rejected only for a very
limited number of subject-categories (see Tables in Supporting Information S9).
For papers published in $1980$, $37\%$ of the subject-categories
have distributions consistent with
log-normals (at $5\%$ significance level).
This proportion, however, decreases for more recent 
years of publication: $28\%$ in $1985$, $20\%$ in $1990$, $10\%$ in $1995$,
$5\%$ in $1999$ and $4\%$ in $2004$. 
While the number of citations received
by papers published in the same year and journal
are log-normally distributed~\cite{stringer08, stringer10},
we should not expect the same for subject-categories.
Subject-categories are given by the aggregation of more journals, and
the convolution of many log-normals with different averages
and variances is not necessarily a log-normal distribution.

\

\noindent
We believe that the methods and results reported in this paper
can be of great relevance for the entire scientific community.
Citation counts and measures based on citations are powerful
tools for the quantitative assessment of science,
especially in our modern era in which millions
of individuals are involved in research but decisions 
(i.e., allocation of funds) need to be quickly taken. 
The use of citations is already a common practice, 
and in the near future will become a necessity. As individuals
directly involved in this business, we should therefore
develop the best methodologies able to
avoid the misuse of citation numbers.

\section{Materials and Methods}

\subsection{Datasets}
\noindent We considered papers published in six distinct years:
$1980$, $1985$, $1990$, $1995$, $1999$ and $2004$. We downloaded from the WOS
database~\cite{wos}  a total of
 $3,964,670$ documents published in $8,304$ scientific journals.
Journal titles have been obtained from~\cite{jcr2009}, and
correspond to all journals classified in 
at least one subject-category by the $2009$ edition of JCR.
According to the JCR classification, a journal may be classified 
in more than one subject-category. For example, the journal
Physical Review D
is classified in the subject-categories ``Astronomy \& astrophysics''
and ``Physics, particles \& fields''.
It is also important to stress that JCR 
classification is made at journal level,
and thus does not allow a
proper distinction of papers in research topics, 
whenever papers are published in multi-category journals. 
In this respect, we adopted, for simplicity, 
a multiplicative strategy, in which
papers published in multi-category journals are
simultaneously associated with all corresponding
subject-categories.
We considered only documents written in ``English'', and
classified as ``Article'', ``Letter'', ``Note'' or
``Proceedings Paper''. 
We obtained a total of
$2,906,615$  publications on which we based our study.
More in detail, we considered in our study
$249,848$ documents published in $1980$, 
$323,296$ in $1985$, $416,378$ in $1990$,
$545,954$ in $1995$, $622,891$ in $1999$ and
$748,248$ in $2004$.
Summary tables regarding the proportion of documents
written in different languages and about
the types of published material can be found
in the Supporting Information S1.
We included in our analysis both cited and 
uncited publications. The information about the 
number of cites received by each publication was obtained
from the WOS database (field ``time cited'') 
between May $23$ and May $31$, $2011$.

\subsection{Test of bias suppression}
\noindent The statistical test proposed here is very similar
to the one introduced in~\cite{radicchi12}.
The unbiased selection of papers
is equivalent to a simple urn model~\cite{mahmoud08},
where papers (marbles) of different subject-categories (colors)
are randomly extracted, one by one, without
replacement. The total number of papers in the urn
is $N$, each subject-category $g$ is represented by
$N_g$ papers, and the total number of
extracted papers is $q = \lfloor N\,v/100 \rfloor$.
The number $m_g$ of papers of subject-category $g$, extracted
in the unbiased selection process, is a random variate
obeying a univariate hypergeometric distribution.
The proportion of papers of subject-category $g$ is still distributed
in the same way, with the only difference of
the change of variable $m_g \to m_g/N_g$ [if 
$P\left(m_g\left|N,q,N_g\right.\right)$
indicates the hypergeometric distribution, the fraction
$m_g/N_g$ obeys the distribution 
$N_g \, P\left(m_g/N_g\left|N,q,N_g\right.\right)$].
Similarly, the joint distribution of the number
of papers $m_{1}$, $m_{2}$, \ldots, $m_{G}$,
belonging respectively to subject-categories 
$1$, $2$, \ldots, $G$ and that have been extracted
in the unbiased selection,
obey a multivariate hypergeometric distribution.
In principle, one could calculate the expected distribution
for the proportions of papers belonging to each subject-category
and that are part of the top $v\%$,
namely $x_v$
by considering
all possible extractions $\{m_1, m_2, \ldots, m_G\}$,
weighting each extraction with the multivariate hypergeometric distribution,
and counting how many times in each extraction
the quantity $m_g/N_g$ (for all subject-categories $g$)
equals $x_v$. 
In practice, it is much simpler to 
simulate many times ($10^4$ times in our analysis)
the process of unbiased selection, and
obtain a good approximation of the
probability density of the proportions of papers
present in the top $v\%$.
This probability
density represents the correct term of comparison
for what observed in real data, and furnishes 
a quantitative criterion for the assessment of 
whether the transformation of Eq.~\ref{eq2} is able to 
suppress subject-category biases in citation counts or not.

\subsection{Supporting Information}
\noindent Supporting Information documents are 
made available at
{\tt http://filrad.homelinux.org}.

\end{document}